\documentstyle[preprint,aps]{revtex}
\tightenlines

\begin{document}
%\articletitle{Quantum Key Distribution with Continuous Variables in 
%Optics}
	\title{Quantum Key Distribution with Continuous Variables in 
Optics}
\author{T.C.Ralph}
\address{Department of Physics, Centre for Laser
Science, \\ University of Queensland, \\ St Lucia, QLD 4072 Australia \\ 
Fax: +61 7 3365 1242  Telephone: +61 7 3365 3412 \\ E-mail: 
ralph@physics.uq.edu.au}
\maketitle

%\articlesubtitle{This is an Article Subtitle}

%\author{T.C.Ralph}
%\affil{Department of Physics\\
%University of Queensland, St Lucia 4072, QLD, Australia}
%\email{ralph@physics.uq.edu.au}

\begin{abstract}
We discuss a quantum key distribution scheme in which small phase and 
amplitude modulations of quantum limited, CW light beams 
carry the key information. We identify universal constraints on 
the level of shared information between the intended receiver (Bob) 
and any eavesdropper (Eve) and use this to make a general evaluation of 
the security and efficiency of the scheme.
\end{abstract}

%\begin{keywords}
%quantum cryptography, quantum information, quantum optics
%\end{keywords}

\section*{Introduction}

%As has been discussed in chapter ? 
The distribution of random number 
keys for cryptographic purposes can 
be made secure by using the fundamental properties of quantum 
mechanics to ensure that any interception of the key information can 
be detected. This was first discussed for discrete systems 
in Refs. \cite{wie83,ben92,eka92}. Experimental demonstrations have 
been carried out using low photon number, optical sources 
\cite{butt98,zbi98}.

The basic mechanism used in quantum cryptographic schemes is the fact 
that the act of measurement in quantum mechanics inevitably 
disturbs the system. This measurement back-action exists for both 
discrete and continuous quantum mechanical variables. Thus it is 
natural to ask if quantum cryptographic schemes based on continuous 
variables are possible. There are a number of practical 
disadvantages with discrete quantum cryptographic schemes, mainly associated 
with the lack of true single photon sources. Also it is of 
fundamental interest to quantum information research to investigate 
links between discrete variable, single photon phenomena and 
continuous variable, multi-photon effects. This has motivated a 
consideration of quantum cryptographic schemes using multi-photon light modes
\cite{ral99,hil00,rei00,kol00,ral00,got00,cer01,gro01}.

Most of these schemes use squeezed light \cite{wal94} in their 
protocols, either by producing entanglement from the squeezing 
\cite{ral00,rei00,kol00} or using the squeezing directly 
\cite{hil00,got00}. In contrast to these, one scheme discussed in 
Ref. \cite{ral99} is based on a coherent state. The signals from which 
the key material is obtained are encoded in various ways in the 
different schemes.
 
The question of optimum protocols and 
eavesdropper strategies has been studied in detail for 
the single quanta case \cite{Fus96}, leading to general proofs 
of security for discrete systems \cite{may96,lo99}. A general proof of the 
optimum eavesdropper strategy for a simple continuous 
variable scheme was presented in Ref. \cite{ral00}. A general proof of 
absolute security for a more sophisticated scheme was presented in Ref.
\cite{got00}.

In this chapter we will analyse in some detail quantum key 
distribution protocols based on the optical coherent state and squeezed state 
schemes introduced in Refs. \cite{ral99,ral00}. Our emphasis will be 
on specific implementations that Alice and Bob might use 
rather than general limits. The particular implementations have been 
chosen mostly for their simplicity rather than their optimality. 
Eve on the other hand is always 
assumed to be employing the optimum eavesdropping strategies allowed by 
quantum physics \cite{note1}. We estimate the efficiency of the two 
schemes and hence secure key transmission rates under conditions of 
negligible and non-negligible losses.

In Section I we review the encoding of information on light with 
small amplitude and phase modulations and introduce a particular 
encoding scheme. In Section II 
we find the minimum disturbance that an 
optimum eavesdropping scheme will introduce. The coherent state 
cryptographic scheme is introduced in Section III and the minimum error 
rates that an optimum Eve will introduce are calculated. In Section 
IV 
the concepts of mutual information, data reconciliation and privacy 
amplification are introduced and specific examples are applied to the 
coherent state scheme. The security and efficiency of the scheme are 
evaluated. The squeezed state cryptographic scheme is introduced, 
analysed and evaluated in Section V. In section VI we discuss a 
physical implementation of the optimal eavesdropper strategy 
and we conclude in Section VII.

\section{Encoding Information with Small Amplitude and Phase Modulations}

One way of encoding information on a light beam is by imposing 
small modulations of the phase or amplitude of the beam at some radio 
frequency (rf) with respect to the main optical frequency. We suppose 
that these signals are imposed at an rf sufficiently large that 
technical noise can be ignored and so our measurement precision is 
limited only by quantum noise. Typically frequencies in excess of 
about a MHz will suffice. That quantum mechanics must impose 
limits in this situation is because the amplitude and and phase quadrature 
amplitudes of the beam are the analogues of position and momentum 
variables. Hence they are continuous, non-commuting variables that exhibit 
uncertainty relations. 

We can represent our light field via
\begin{equation}
\hat a(t)=\alpha + \delta \hat a(t) + \delta s(t)
\end{equation}
where $\hat a$ is a bosonic annihilation operator which we have decomposed 
into a steady state part, the coherent amplitude, $\alpha$, treated 
classically, and two time varying parts: the quantum fluctuations, 
modelled by the operator $\delta \hat a(t)$; and the classical 
modulation, modelled by $\delta s(t)$. If we take 
the phase of $\alpha$ real then the amplitude fluctuations, $\hat 
X^{+}$, and the phase fluctuations, $\hat X^{-}$, are given by 
\begin{eqnarray}
\hat X^{+} & = & \delta \hat a(t)^{\dagger} + \delta s(t)^{*}
+\delta \hat a(t) + \delta s(t) \nonumber\\
\hat X^{-} & = & i(\delta \hat a(t)^{\dagger} + \delta s(t)^{*}
-\delta \hat a(t) - \delta s(t))
\end{eqnarray}
Homodyne detection using a 
local oscillator with a coherent amplitude much larger than that of 
the signal beam can be used to measure the fluctuations. Spectral 
analysis then extracts the fluctuation power at a particular rf, 
$\omega$, such 
that
\begin{equation}
V^{+}(\omega)=<|\tilde X^{+}|^{2}>=V_{n}^{+}+V_{s}^{+}
\end{equation}
and
\begin{equation}
V^{-}(\omega)=<|\tilde X^{-}|^{2}>=V_{n}^{-}+V_{s}^{-}
\end{equation}
where $V_{n}^{+}$ ($V_{n}^{-}$) is the amplitude (phase) quantum noise 
power whilst  $V_{s}^{+}$ ($V_{s}^{-}$) is the amplitude (phase) 
signal power. The tilde indicates a Fourier transform.

The amount of information that can be carried on a Gaussian, additive noise, 
communication channel, such as we will consider here, 
depends on the signal to noise \cite{har49}. For a fixed bandwidth, 
any reduction in the signal to noise will inevitably lead to 
increased errors in the transmission. In our cryptographic scheme 
signals will be encoded on both 
quadratures but read out from only one, randomly chosen. This will 
force any eavesdroppers to monitor both the amplitude and phase quadratures 
simultaneously. For these non-commuting observables 
the information that can be obtained in this way is 
strictly limited by the generalized uncertainty principle for 
simultaneous measurements \cite{yam86,aut88}. We will discuss this 
principle in detail in the next section. 
Here let us consider a 
simple example.  Suppose we 
try to observe both quadratures by dividing the beam in two at a 
50:50 beamsplitter and detecting the amplitude quadrature of one beam 
and the phase quadrature of the other. Originally the signal to 
noises are given by
\begin{equation}
(S/N)^{\pm}={{V_{s}^{\pm}}\over{V_{n}^{\pm}}}
\end{equation}
However the signal to noises detected after the beamsplitter are
\begin{equation}
(S/N)_{sim}^{\pm}=({{V_{n}^{\pm}}\over{
V_{n}^{\pm}+V_{m}^{\pm}}}) S/N^{\pm}= T^{\pm}S/N^{\pm}
\label{sn}
\end{equation}
where we define $T^{+}$ ($T^{-}$), the amplitude (phase) signal transfer 
coefficient, as the ratio of signal to noise out to signal to noise in. 
The quantum noise which is inevitably added through the 
empty beamsplitter port is $V_{m}^{\pm}$. The spectral powers are 
normalized to the quantum 
noise limit (QNL) such that a coherent beam has $V_{n}^{\pm}=1$. Normally 
the partition noise will also be at this limit ($V_{m}^{\pm}=1$). For 
a classical light field, i.e. where $V_{n}^{\pm}>>1$ the penalty will 
be negligible. However for a coherent beam a halving of the signal to 
noise for both quadratures is unavoidable.

To continue our analysis of this example let us consider the specific 
encoding scheme of binary pulse code modulation. The data is 
encoded as a train of pulses, a pulse on representing a ``1'', a 
pulse off representing a ``0''. For bandwidth limited transmission
the bit error rate or error probability 
(${\cal B}$) and the signal to noise ($S/N$) are 
related by \cite{yar97}
\begin{equation}
{\cal B}={{1}\over{2}}erfc{{1}\over{2}}\sqrt{{{1}\over{2}}S/N}
\label{ber}
\end{equation}
Now suppose our signal to noise is initially 13 dB. From Eq.\ref{ber} 
direct detection of a single quadrature will retrieve its pulse train 
with a bit error rate of 1\%. If the beam is in a coherent state and 
we simultaneously detect both quadratures then 
Eq.\ref{sn} tells us that the signal to noise is halved. 
Eq.7 then predicts the error rate will rise to 5\%. 

\section{Optimum Eavesdropper Strategy}

In this section we will use the generalized uncertainty principle 
to identify the minimum 
disturbance allowed by quantum mechanics to the information Bob 
receives given a particular level of interception by Eve. The idea is 
shown schematically in Fig.1. A single quantum limited beam is sent 
from Alice to Bob. Eve makes some unspecified interception of the 
beam enroute. Bob and Eve obtain some measurement results. We will 
show that quantum mechanics sets unambiguous limits on the level of 
quantum noise that must appear in Bob and Eve's results. In the 
following sections we will apply these results to specific quantum 
cryptographic systems.

A more general statement of the generalized uncertainty 
principle \cite{aut88} requires that for 
{\it any} simultaneous measurements of conjugate quadrature amplitudes
\begin{equation}
V_{M}^{+} V_{M}^{-}\ge 1
\label{gu}
\end{equation}
where $V_{M}^{\pm}$ are the measurement penalties for the 
amplitude ($+$) and phase ($-$) quadratures, normalized to the 
amplification gain between the system observables and the measuring 
apparatus. For example suppose an attempt to measure the amplitude 
quadrature variance of a system $V_{k}^{+}$ returned the result $G_{1} 
V_{k}^{+}+G_{2}V_{m}^{+}$ where $V_{m}^{+}$ represents noise. Then we 
would have $V_{M}^{+}=(G_{2}/G_{1})V_{m}^{+}$. Eq.\ref{sn} follows 
directly from  Eq.\ref{gu} for ideal simultaneous measurements. Let 
us investigate what general restrictions this places on the 
information that Eve can intercept and the subsequent corruption of 
Bob's signal. Firstly Eve's measurements will inevitably carry 
measurement penalties $V_{E}^{\pm}$ constrained by 
\begin{equation}
V_{E}^{+} V_{E}^{-}\ge 1
\label{eve}
\end{equation}
Now suppose Bob makes an ideal (no noise added) 
amplitude measurement 
on the beam he receives.  In order to satisfy Eq.\ref{gu} 
it must be true that the noise penalty 
carried on the amplitude quadrature of this beam $V_{B}^{+}$ due to 
Eve's intervention, 
is sufficiently large such that 
\begin{equation}
V_{B}^{+} V_{E}^{-}\ge 1
\label{bp}
\end{equation}
Similarly, 
Bob can also choose to make ideal measurements of the phase quadrature so we must 
also have 
\begin{equation}
V_{E}^{+} V_{B}^{-}\ge 1
\label{bm}
\end{equation}
Eqs.\ref{eve},\ref{bp},\ref{bm} set strict quantum mechanical limits on the 
minimum disturbance Eve can cause to Bob's information given a 
particular maximum quality of the information she receives. This 
applies regardless of the method she uses to eavesdrop. Note that 
quantum memory does not negate the above results provided we insist 
that Alice and Bob do not exchange any potentially revealing classical 
information until Alice is sure that Bob has received and measured her 
signals. 

These relations could form the basis of a security analysis of any 
continuous variable quantum cryptographic scheme in which a single 
quantum beam is exchanged. However the ramifications of a particular 
level of disturbance will vary for different schemes. In the 
following section we will analyse the security of a very simple 
scheme based on the exchange of a beam in a coherent state.

\section{Coherent State Quantum Cryptography}

Consider the set up depicted in Fig.2. A possible protocol is 
as follows. Alice generates two 
independent random strings of numbers and encodes one on the phase 
quadrature, and the other on the amplitude 
quadrature of a bright coherent beam. The amplitude and phase signals are 
imposed at the same frequency with 
equal power. Bob uses homodyne detection to 
detect either the amplitude or phase quadrature of the beam when he 
receives it. He swaps randomly which quadrature he detects. On a 
public line Bob then tells Alice at which quadrature he was looking, at 
any particular time. They pick some subset of Bob's data to be the test and 
the rest to be the key. For example, they may pick the amplitude 
quadrature as the test signal. They would then compare results for the times that 
Bob was looking at the amplitude quadrature. If Bob's results agreed 
with what Alice sent, to within some acceptable error rate, they would 
consider the transmission secure. They would then use the undisclosed 
phase quadrature signals, sent whilst Bob was observing the phase 
quadrature, to create their key. By randomly swapping which quadrature is key 
and which is test throughout the data comparison an increased error 
rate on either quadrature will immediately be obvious.

Before making a general analysis of security let us first consider 
some specific strategies an eavesdropper could adopt. Eve 
could guess which quadrature Bob is going to measure and measure it 
herself. 
She could then reproduce the digital signal of that 
quadrature and impress it on another coherent beam which she would 
send on to Bob. She would learn nothing about the other quadrature 
through her measurement and would have to guess her own random string 
of numbers to place on it. When Eve guesses the right quadrature to 
measure Bob and Alice will 
be none the wiser, however, on average 50\% of the time Eve will guess 
wrong. Then Bob will receive a random string from Eve unrelated to the 
one sent by Alice. These will agree only 50\% of the time. 
Thus Bob and Alice would see a 25\% bit error rate in the test 
transmission if Eve was using this strategy. This is analogous to the 
result for single quanta schemes in which this type of strategy 
is the most readily available. Another single measurement 
strategy Eve could use is to do homodyne 
detection at a quadrature angle half-way between phase and amplitude. 
This fails because the signals become mixed. Thus Eve can tell when 
both signals are 0 or both are 1 but she cannot tell the 
difference between 1,0 and 0,1. This again leads to a 25\% bit error 
rate.

However, for bright beams it is possible to make simultaneous 
measurements of the quadratures, with the caveat that there will be 
some loss of information. So a second strategy that Eve could follow 
would be to split the beam in half, measure both quadratures and 
impose the information obtained on the respective quadratures of 
another coherent beam which she sends to Bob. 
How well will this 
strategy work? We performed this calculation at the end of section I 
using Eq.\ref{ber}. The halving of signal to noise imposed by the 
50:50 beamsplitter means the information Eve intercepts and subsequently 
passes on to Bob will have an error probability of 5\% 
(for the particular case of 
bandwidth limited binary pulse code modulation). This is clearly a superior 
strategy and would be less easily detected. Further more Eve could 
adopt a third strategy of only intercepting a small amount of the 
beam and doing simultaneous detection on it. For 
example, by intercepting 16\% of the beam, Eve could gain information 
about both quadratures with an error rate of 25\% whilst Bob and Alice would 
observe only a small increase of their error rate to 1.7\%. In other words Eve 
could obtain about the same amount of information about the key that she 
could obtain using the ``guessing'' strategy, whilst being more 
difficult to detect.

Now let us analyze this coherent state scheme using 
Eqs.\ref{eve},\ref{bp},\ref{bm}. We choose to couch our evaluation in 
terms of bit error rates because they represent an unambiguous, directly 
observable measure of the extent to which Eve can intercept 
information and the resulting corruption of Bob's information. This 
connection will be developed in Section IV.
Depending on the particular technique Eve uses Bob and 
Alice may be able to gain additional evidence for Eve's presence by 
comparing the absolute noise levels of the sent and received signals. 
This can only increase the security of the system. By considering a 
general limit on error rates we can find a minimum guaranteed security 
against eavesdropping regardless of the technique Eve employs.

The signal transfer 
coefficients for Bob and Eve will be given by
\begin{eqnarray}
T_{E}^{+} & = & 
{{(S/N)_{eve}^{+}}\over{(S/N)_{in}^{+}}}=
{{V_{in}^{+}}\over{V_{in}^{+}+V_{E}^{+}}}\nonumber\\
T_{E}^{-} & = & 
{{(S/N)_{eve}^{-}}\over{(S/N)_{in}^{-}}}=
{{V_{in}^{-}}\over{V_{in}^{-}+V_{E}^{-}}}\nonumber\\
T_{B}^{+} & = & 
{{(S/N)_{bob}^{+}}\over{(S/N)_{in}^{+}}}=
{{V_{in}^{+}}\over{V_{in}^{+}+V_{B}^{+}}}\nonumber\\
T_{B}^{-} & = & 
{{(S/N)_{bob}^{-}}\over{(S/N)_{in}^{-}}}=
{{V_{in}^{-}}\over{V_{in}^{-}+V_{B}^{-}}}
\label{ts}
\end{eqnarray}
Substituting Eqs.\ref{ts} into Eqs.\ref{eve},\ref{bp},\ref{bm} and 
using the fact that $V_{in}^{\pm}=1$ we find
\begin{eqnarray}
T_{E}^{+}+T_{E}^{-} & \le & 1 \nonumber\\
T_{E}^{+}+T_{B}^{-} & \le & 1 \nonumber\\
T_{B}^{+}+T_{E}^{-} & \le & 1 
\label{ts2}
\end{eqnarray}
Eqs.\ref{ts2} clearly show that any attempt by Eve to get a good 
signal to noise on one quadrature (e.g. $T_{E}^{+}\to 1$) 
results not only in a 
poor signal to noise in her information of the other quadrature 
(e.g. $T_{E}^{-}\to 0$) but 
also a poor signal to noise for Bob on that quadrature (e.g. 
$T_{B}^{-}\to 0$), making her presence obvious. This is the general 
limit of the guessing strategy presented in the last section and leads 
to the same error rates.

Because of the symmetry of Bob's readout technique Eve's best 
approach is a symmetric attack on both quadratures. Eqs.\ref{ts2} then 
reduces to two equations
\begin{eqnarray}
2 T_{E}^{\pm} & \le & 1 \nonumber\\
T_{E}^{\pm}+T_{B}^{\pm} & \le & 1
\label{ts3}
\end{eqnarray}
If Eve extracts her maximum allowable signal to noise transfer, 
$T_{E}^{\pm}=0.5$, then ideally Bob suffers the same penalty 
$T_{B}^{\pm}=0.5$. This is the general limit of the second strategy of the 
previous section. The same reduction in Bob's signal to noise occurs 
as in the specific implementation thus this implementation can be 
identified as an optimum eavesdropper strategy for obtaining maximum 
simultaneous information about both quadratures. 

Eve's best strategy is to intercept only as much information as she 
can without being detected. The system will be secure if that level 
of information can be made negligible. Suppose, as in the last section, Eve 
only intercepts a signal transfer of $T_{E}^{\pm}=.08$. 
From Eq.\ref{ts3} this means Bob can receive at most a signal transfer of 
$T_{B}^{\pm}=.92$. This is greater than the result for the specific 
implementation discussed in the last section, thus that implementation is not  
an optimum eavesdropper strategy. Using the optimum eavesdropper 
strategy the error rates for the specific 
encoding scheme discussed in the last section will be: if Eve 
intercepts information with an error probability of 
25\%, then the minimum error rate in Bob's information will be 1.4\%.

In Fig.3 we represent the general situation by plotting the minimum 
error rate Bob and Alice can observe against the error rate 
in Eve's intercepted information. An error rate of 50\% (i.e. 
completely random) represents 
no information about the data. Two traces are shown, representing 
different initial signal to noises in Alice's data. 
This graph shows that in principle any 
incursion by Eve will result in some increase in Bob and Alice's 
error rate. However one could also argue that any finite 
resolution in Bob and Alice's determination of their error rate will 
allow Eve to do better than the random result. In order to assess 
whether this system can be made secure we need to introduce the 
concepts of mutual information and privacy amplification.

\section{Mutual Information and Privacy Amplification}

The mutual information of party 1 and party 2 is the information overlap between 
the data possessed by the two parties. The binary entropy of party 1's data, 
$x$, is given by
\begin{equation}
	H(x)=-p_{x} log_{2} p_{x} -(1-p_{x}) log_{2}(1-p_{x})
\end{equation}
where $p_{x}$ and $1-p_{x}$ are the probabilities of the two outcomes. 
Similarly party 2's data, $y$, has binary entropy
\begin{equation}
	H(y)=-p_{y} log_{2} p_{y} -(1-p_{y}) log_{2}(1-p_{y})
\end{equation}
The joint entropy of the two data strings is then given by
\begin{equation}
	H(x,y)=-\Sigma_{x,y} p_{x,y} log_{2} p_{x,y}
\end{equation}
with $p_{x,y}$ the joint probabilities. The mutual information is 
defined
\begin{equation}
	H(x:y)=H(x)+H(y)-H(x,y)
\end{equation}
If the two data strings $x$ and $y$ are random then $H(x)=H(y)=1$. 
Suppose the error probability between the data strings is ${\cal B}$, 
then the joint probabilities are 
given by $p_{0,0}=p_{1,1}=1-{\cal B}$ and $p_{0,1}=p_{1,0}={\cal B}$. 
Thus we find
\begin{equation}
	H(x:y)=1+{\cal B} log_{2} {\cal B} +(1-{\cal B}) 
	log_{2}(1-{\cal B})
	\label{mi}
\end{equation}
Suppose $A$ is Alice's data string, $B$ is Bob's data string and $E$ is Eve's 
data string. Maurer has shown \cite{mau93} that provided $H(A:B)>H(A:E)$ 
then it is in principle possible for Alice and Bob to extract a secret 
key from the data. Eve's mutual information with this secret key can be made 
arbitrarily small. From Eq.\ref{mi} we see that this condition will be 
satisfied provided Bob's error rate is less than Eve's. From Fig.3 we 
see that provided Alice and Bob's error rate does not exceed 5\% for 
case (a) or 12\% for case (b) then secret key generation is in 
principle possible. In the following we will look at a simple specific 
example of a secret key generation protocol and evaluate its 
efficiency.

Because of the transmission errors (and possibly the actions of Eve) 
Alice and Bob won't share the same data string. However 
techniques exist for data reconciliation which allow Alice and Bob to 
select with high probability a subset of their data which is error 
free, whilst giving Eve minimal extra knowledge. As a simple example 
Alice and Bob could perform a parity check on randomly chosen pairs of 
bits. If the error rate between Bob and Alice is low then the probability of 
both bits being wrong is very low. Thus discarding all pairs which 
fail the parity check will lead to a big reduction in errors 
in the shared data whilst not revealing the values of the individual 
bits to Eve. A series of parity checks will lead with high probability to 
zero errors. Eve can also remove the pairs that Bob removes and in a 
worse case scenario may remove up to the same number of errors as 
Bob. But if Eve initially had significantly more errors than Bob then 
she will still have significant errors after the reconciliation, 
whilst Bob and Alice will have virtually none. The data string length 
will be reduced by a factor of approximately $1-2 {\cal B}_{B}$, 
where ${\cal B}_{B}$ is Bob's error probability.

In order to reduce Eve's mutual information to a negligible amount 
the technique of privacy amplification is employed \cite{ben95}. 
This involves the 
random hashing or block coding of the reconciled key into a shorter key. 
As a simple example Alice and Bob could randomly 
pick data strings of length $n$ from 
the reconciled key and form a new key from the sum, modulo 2, of 
each $n$ unit block. It is important that the privacy amplification 
is ``orthogonal'' to the reconciliation protocol. That is none of the 
pairs used in the parity checks should appear together in the privacy 
amplification blocks. The length of the new key will be reduced by a factor 
of $1/n$. The error probability in the new key will be given by
\begin{equation}
	{\cal B}_{pa}=\Sigma_{k=0}^{n/2} {{n!}\over{(2 k)!(n-2 k)!}} (1-{\cal 
	B})^{n-2 k} 
	{\cal B}^{2 k}
	\label{pa}
\end{equation}
where ${\cal B}$ is the error probability of the original string. If 
${\cal B} \approx 
0$, as for Bob and Alice, then this process introduces virtually no 
errors. But when Eve copies this process her errors will be 
``amplified'', hopefully to the point where her mutual information is 
negligible. Some caution is required in evaluating Eve's mutual information 
now. Just as Bob and Alice were able to select a sub-set of results 
they knew were correct in the reconciliation process, so Eve can also obtain 
a (smaller) subset of results for which she has greater confidence. 
We make the worst case assumption that after privacy amplification Eve 
is left with some small probability, $p_{r}$, of possessing certain bits that 
she knows are right, and a large probability, $1-p_{r}$, of 
possessing bits which are completely random. In such a situation it is 
appropriate to set Eve's mutual information as
\begin{equation}
	H(A:E)=p_{r}=1-2 {\cal B}_{pae}
	\label{he}
\end{equation}
where ${\cal B}_{pae}$ is Eve's average error probability, as given by 
Eq.\ref{pa}.

Let us now apply these techniques to the continuous variable protocol 
of the previous section to evaluate its security. After Bob has 
received all the data from Alice he tells her at which quadrature he was 
looking at any particular time and Alice sorts out her sent data 
accordingly. They then compare a randomly chosen sub-section of their 
data (approximately half) and determine the error rate. For the 
example in the previous section they expect a base error rate of 1\%. 
Let them reject the data and start again if they detect an error rate 
of $\ge$ 2\%. To be cautious, let us assume that in fact the error rate 
could have been as high as 2.5\%. For sufficiently long data strings 
there will be negligible probability of this error rate being exceeded 
in the undisclosed data \cite{nei00}. From Fig.3 (trace (a)) we can read off that 
Eve's error rate must be $\ge$10.5\%. Applying our simple information 
reconciliation protocol Bob and Alice's error probability 
can be reduced to virtually zero whilst Eve's error rate is $\ge$8\%. 
We now apply 
privacy amplification. Fig.4 (trace (a)) shows Eve's mutual information as a 
function of the block length, $n$. Clearly Eve's mutual information is 
decreasing exponentially as a function of block length. This is the 
signature of a secure system. A linear expenditure of resources 
results in an exponentially small mutual information with Eve. In fact 
Bob and Alice can do better by using a smaller initial signal 
strength. If Alice reduces the size of the signal she sends to about 
half that of the previous example (now with a signal to noise of 
about 10dB) Bob's base error rate will rise to 5\%. They set their 
error threshold at 6\%. To be cautious we assume the error rate could 
be as high as 6.5\%. From Fig.3 (trace (b)) we find that Eve's 
error rate must be $\ge$26\%. After reconciliation Eve's error rate must 
still be $\ge$19.5\%. Fig.4 (trace(b)) plots Eve's mutual information as a 
function of the block length for this situation showing a more rapid 
decay. This is approximately the optimum signal strength. However Alice 
and Bob may also seek to improve the efficiency of the system by 
employing more sophisticated 
reconciliation and privacy amplification protocols.

To this point we have assumed that the transmission line between Alice 
and Bob is lossless. In practice this will not be true. If we make no 
constraints on Eve's technical abilities then we must assume that all 
lost light has fallen into her hands \cite{note2}. Thus we must calculate Eve's 
potential mutual information from Bob's error rate as if there was no 
loss, but we must set our error threshold quite high because the 
losses will drive up Bob's errors. It is clear that loss of 50\% or 
more can not be tolerated because Eve's and Bob's error rates become equal at 
this point. Indeed as losses approach 50\% the expenditure of resources 
by Bob and Alice needed to 
reconcile and privacy amplify will increase rapidly.

Let us estimate by what factor the length of the final secure key would be 
reduced over the length of the original string sent by Alice in a system 
with 25\% loss. 
Consider an original signal to noise 
of about 10dB, leading to a base error rate with 25\% loss of 7.7\%. 
Setting as before our maximum error rate 1.5\% above the base rate at 9.3\% we 
can bound Eve's error rate at $\ge$16.3\%. Bob and Alice sacrifice half 
their data in this step. Reconciliation will reduce Bob and Alice's 
data string by a factor of 0.81 and leave Eve with an error rate $\ge$7\%. 
If we require that Eve's mutual information be $\le$0.001 for the 
transmission to be considered secure then we find a block length of 
$n=46$ is required in the privacy amplification step. Thus the secure 
key will be reduced by a factor of $0.5 \times 0.81 \times 
0.02=0.01$. (A similar estimate for the optimum no loss case 
gives a reduction factor of 0.025) 
Data transmission via rf signals is a mature technology and bit 
transmission rates of 100 MHz would seem quite reasonable. Thus secure key 
transmission rates of a a MHz would seem practical under these 
conditions. This is about three orders of magnitude better than what 
is presently achievable with single photon schemes. On the 
other hand it should be 
noted that single quanta schemes can tolerate much higher losses 
\cite{butt98}. 

\section{Squeezed State Quantum Cryptography}

The preceding discussion has shown that a cryptographic scheme based 
on coherent light can produce secure keys with an efficiency of about 
$1/40 \to 1/100$. We now consider whether squeezed light can offer improved 
efficiency.

The set-up is shown in Fig.5. Once again Alice encodes her number 
strings digitally, but now she impresses them on the amplitude 
quadratures of two, phase locked, 
amplitude squeezed beams, $a$ and $b$, one on each. A $\pi/2$ phase 
shift is imposed on beam $b$ and then they are mixed on a 
50:50 beamsplitter. The resulting output modes, $c$ and $d$, are given by
\begin{eqnarray}
c & = & \sqrt{{{1}\over{2}}}(a+i b)\nonumber\\
d & = & \sqrt{{{1}\over{2}}}(a-i b)
\end{eqnarray}
These beams are now in an entangled state which will exhibit Einstein, 
Podolsky, Rosen (EPR) type correlations \cite{ein35,ral98}. Negligible 
information about the  signals can be extracted from the beams 
individually because the large fluctuations of the anti-squeezed 
quadratures are now mixed with the signal carrying squeezed 
quadratures. One of the beams, say $c$, is 
transmitted to Bob. The other beam, $d$, 
Alice retains and uses homodyne detection 
to measure either its amplitude or phase fluctuations, with respect to a 
local oscillator in phase with the original beams $a$ and $b$. She 
randomly swaps which quadrature she measures, and stores the results. 
Bob, upon receiving beam $c$, also randomly chooses to measure either 
its amplitude or phase quadrature and stores his results. After the 
transmission is complete Alice sends the results of her measurements 
on beam $d$ to Bob on an open channel. About half the time Alice 
will have measured a different quadrature to Bob in a particular time window. 
Bob discards these results. The rest of the data corresponds to times 
when they both measured the same quadratures. If they both measured 
the amplitude quadratures of each beam Bob adds them together, in which case
he can obtain the power spectrum
\begin{eqnarray}
V^{+} & = & <|(\tilde c^{\dagger}+\tilde c)+(\tilde d^{\dagger}
+\tilde d)|^{2}>\nonumber\\
 & = & V_{s,a}+V_{n,a}^{+}
\end{eqnarray}
where the tilde indicate Fourier transforms. Thus he obtains the data string 
impressed on beam $a$, $V_{s,a}$, 
imposed on the sub-QNL noise floor of beam $a$, $V_{n,a}^{+}$. 
Alternatively if they both measured the phase quadratures of each 
beam, Bob subtracts them, in which case he can obtain the power spectrum
\begin{eqnarray}
V^{-} & = & <|(\tilde c^{\dagger}-\tilde c)-(\tilde 
d^{\dagger}-\tilde d)|^{2}>\nonumber\\
 & = & V_{s,b}+V_{n,b}^{+}
\end{eqnarray}
i.e. he obtains the data string impressed on beam $b$, $V_{s,b}$, 
imposed on the sub-QNL noise floor of beam $b$, $V_{n,b}^{+}$. Thus 
the signals lie on conjugate quadratures but 
{\it both} have sub-QNL noise floors. This is the hallmark of the EPR 
correlation \cite{ou92}. As for the coherent state case Alice and Bob 
now compare some sub-set of their shared data and check for errors. If 
the error rate is sufficiently low they deem their transmission 
secure and use reconciliation and privacy amplification on the undisclosed 
sub-set of their data to produce a secure key.

Consider now eavesdropper strategies. Eve must intercept beam $c$ if 
she is to extract any useful information about the signals from the classical 
channel (containing Alice's measurements of beam $d$) sent later. 
She can adopt the guessing strategy by 
detecting a particular 
quadrature of beam $c$ and then using a similar apparatus to Alice's 
to re-send the beam and a corresponding classical channel later. 
As before she will only guess correctly what Bob will measure half the 
time thus introducing a BER of 25\%. Instead she may try 
simultaneous detection of both quadratures of beam $c$. As in the 
coherent case the noise she introduces into her own measurement 
($V_{E}^{\pm}$) and that she introduces into Bob's ($V_{B}^{\pm}$) are 
in general limited according to Eqs.\ref{eve},\ref{bp} and \ref{bm}. 
However now the consequences of these noise limits on the signal to 
noise transfers that Eve and Bob can obtain behave quite differently 
because the signals they are trying to extract lie on sub-QNL 
backgrounds. The maximum signal transfer coefficients that Eve can 
extract are given by
\begin{eqnarray}
T_{E}^{+} & = & {{(V_{E}^{+}+2 V_{n,b}^{-}) 
V_{n,a}^{+}}\over{2 V_{n,a}^{+} V_{n,b}^{-}+
 V_{E}^{+}(V_{n,a}^{+}+V_{n,b}^{-})}}\nonumber\\
T_{E}^{-} & = & {{(V_{E}^{-}+2 V_{n,a}^{-}) 
V_{n,b}^{+}}\over{2 V_{n,b}^{+} V_{n,a}^{-}+
 V_{E}^{-}(V_{n,b}^{+}+V_{n,a}^{-})}}
\label{seve0}
\end{eqnarray}
Similarly Bob's are
\begin{eqnarray}
T_{B}^{+} & = & {{(V_{B}^{+}+2 V_{n,b}^{-}) 
V_{n,a}^{+}}\over{2 V_{n,a}^{+} V_{n,b}^{-}+
 V_{B}^{+}(V_{n,a}^{+}+V_{n,b}^{-})}}\nonumber\\
T_{B}^{-} & = & {{(V_{B}^{-}+2 V_{n,a}^{-}) 
V_{n,b}^{+}}\over{2 V_{n,b}^{+} V_{n,a}^{-}+
 V_{B}^{-}(V_{n,b}^{+}+V_{n,a}^{-})}}
\label{sb0}
\end{eqnarray}
To achieve maximum security we require that the anti-squeezed 
quadratures of the beams have large excess noise. This could easily be
arranged experimentally. The maximum signal transfer coefficients 
(Eq.\ref{seve0} and Eq.\ref{sb0}) then reduce to 
\begin{eqnarray}
T_{E}^{+} & = & {{V_{n,a}^{+}}\over{V_{n,a}^{+}+0.5 V_{E}^{+}}}\nonumber\\
T_{E}^{-} & = & {{V_{n,b}^{+}}\over{V_{n,b}^{+}+0.5 V_{E}^{-}}}
\label{seve}
\end{eqnarray}
and similarly Bob's are
\begin{eqnarray}
T_{B}^{+} & = & {{V_{n,a}^{+}}\over{V_{n,a}^{+}+0.5 V_{B}^{+}}}\nonumber\\
T_{B}^{-} & = & {{V_{n,b}^{+}}\over{V_{n,b}^{+}+0.5 V_{B}^{-}}}
\label{sb}
\end{eqnarray}
For the squeezed noise floors the same ($V_{n,a}^{+}=V_{n,b}^{+}=V_{n}$) 
we find the signal transfers are restricted via
\begin{equation}
4 V_{n}^{2}({{1}\over{T_{E}^{+}}}-1)({{1}\over{T_{E}^{-}}}-1)\ge 1
\label{te}
\end{equation}
\begin{equation}
4 V_{n}^{2}({{1}\over{T_{E}^{+}}}-1)({{1}\over{T_{B}^{-}}}-1)\ge 1
\label{tb}
\end{equation}
\begin{equation}
4 V_{n}^{2}({{1}\over{T_{B}^{+}}}-1)({{1}\over{T_{E}^{-}}}-1)\ge 1
\label{tbb}
\end{equation}
It is straightforward to show that a 
symmetric attack on both quadratures is Eve's best strategy as it 
leads to a minimum disturbance in both her and Bob's measurements. 
Using this symmetry to simplify Eq.\ref{te} 
leads to the following general restriction on the 
signal transfer Eve can obtain:
\begin{equation}
T_{E}^{\pm} \le {{2 V_{n}}\over{2 V_{n}+1}}
\label{tes}
\end{equation}
Once the squeezing exceeds 3 dB ($V_{n}=0.5$) the signal to noise that 
Eve can obtain simultaneously is reduced below that for the coherent 
state scheme. In the limit of very strong squeezing ($V_{n}\to 0$) Eve 
can extract virtually no information simultaneously. Similarly Bob's signal 
transfer is restricted according to:
\begin{equation}
{{T_{E}^{\pm} T_{B}^{\pm}}\over{(1-T_{E}^{\pm})(1-T_{B}^{\pm})}} 
\le 4 V_{n}
\label{tbs}
\end{equation}
If squeezing is strong then almost any level of interception by Eve 
will result in very poor signal transfer to Bob. In Fig.6 we show 
plots of error rates of Bob versus minimum error rates of Eve for 
various levels of squeezing. In comparison with the coherent scheme 
(Fig.3) it can be seen 
that larger disturbances are caused in Bob's information for the same 
quality of Eve's interception. As a numerical 
example consider the specific encoding scheme of section I and suppose 
the squeezing is 10 dB ($V_{n}=0.1$). Assuming no loss and using the 
same assumptions as those used to evaluate the coherent scheme 
in the last section we find that a secure key of length $0.07$ times 
the original data string length can be generated. That is an 
efficiency of about $1/14$, to be compared to the coherent case of 
$1/40$, a clear improvement. 

Unfortunately this high sensitivity to interception by Eve also 
results in a high sensitivity to loss. For the case above the break 
even point between Bob and Eve's errors is for a loss of only 16\%. 
The squeezing system can only be used to advantage if the loss in the 
transmission is much less than this value. Thus we find that although 
in principle squeezing improves the efficiency of the scheme, in 
practice the constraints on the quality of the transmission become 
quite critical. 

\section{Teleportation as the Optimum Eavesdropper Strategy}

It is interesting to consider what physical techniques Eve could use 
to realize the optimal attack strategy we have assumed her capable of 
throughout this discussion. Firstly she would need to replace the lossy 
transmission line that Bob and Alice are using with her own 
transmission line of negligible loss. Given that Bob and Alice will 
presumably employ the most efficient transmission line they can 
obtain, Eve's job is not trivial. Secondly she must extract 
information from the transmission with the least possible 
disturbance, such that the inequalities of Eqs.\ref{eve},\ref{bp} and 
\ref{bm} are 
saturated. In Ref.\cite{ral00} it was shown that Eve can use continuous variable 
teleportation \cite{vai94,bra98,ral98} as such an optimum eavesdropper 
strategy. In this section we will review that result.

Quantum teleportation uses shared entanglement to convert 
quantum information into classical information and then back again 
(see Fig.6). In 
particular continuous variable teleportation uses 2-mode squeezed light 
as its entanglement resource. In the limit of very strong squeezing no 
information about the teleported system can be extracted from the 
classical channel but a perfect reproduction of the quantum system 
can be retrieved. On the other hand with lower levels of squeezing 
some information about the system can be obtained from the the 
classical channel but at the expense of a less than perfect 
reproduction. We show in the following that under particular operating 
conditions the disturbance in the teleported state is precisely the 
minimum required by the generalized uncertainty principle, given the 
quality of information that can be extracted from the classical channel. 
Teleportation thus constitutes an optimum eavesdropper strategy.

Eve's strategy would be to send the field she intercepts from Alice 
through a teleporter, adjusted such that she can read some 
information out of the classical channel, but still reconstruct the 
field sufficiently well such that Bob and Alice don't see a large 
error rate.
The classical channel of a lossless continuous variable teleporter can 
be written \cite{ral00,ral98}
\begin{eqnarray}
F_{c} & = & K(\hat f_{in}+\hat j_{1}^{\dagger})\nonumber\\
 & = & K(\hat f_{in}+\sqrt{G}\hat v_{1}^{\dagger}+\sqrt{G-1}\hat v_{2})
\end{eqnarray}
where $\hat f_{in}$ is the annihilation operator of the input to the 
teleporter and $\hat j_{1}=\sqrt{G}\hat v_{1}+\sqrt{G-1}\hat v_{2}^{\dagger}$ 
is the 
annihilation operator for one of the entangled beams. The $\hat v_{i}$ are 
the vacuum mode inputs to the squeezers, $G$ is the parametric gain of 
the squeezers and $K>>1$ is the measurement amplification 
factor. Being a classical channel simultaneous measurements of both 
quadratures can be made without additional penalty thus immediately 
Eve's measurement penalty is
\begin{equation}
V_{E}^{\pm}=2 G-1
\end{equation}
For no squeezing ($G=1$) $V_{E}^{\pm}=1$, the minimum possible for 
simultaneous detection of both quadratures (see Eq.\ref{eve}). For 
large squeezing ($G>>1$) $V_{E}^{\pm}$ become very large and Eve can 
obtain little 
information from the classical channel.

The output of the teleporter is given by
\begin{eqnarray}
\hat f_{out} & = & \lambda \hat f_{in}+\hat j_{1}^{\dagger}-\hat j_{2}\nonumber\\
 & = & \lambda \hat f_{in}+(\lambda \sqrt{G}-\sqrt{G-1})\hat v_{1}^{\dagger}+
(\sqrt{G}-\lambda \sqrt{G-1})\hat v_{2}
\end{eqnarray}
where $\lambda$ is the gain of the teleporter and $\hat 
j_{2}=\sqrt{G}\hat v_{2}+
\sqrt{G-1}\hat v_{1}^{\dagger}$ is the 
annihilation operator for the other entangled beam. Thus Bob's measurement 
penalty for ideal measurements of either of the quadratures is
\begin{equation}
V_{B}^{\pm}= {{(\lambda \sqrt{G}-\sqrt{G-1})^{2}+(\sqrt{G}-\lambda 
\sqrt{G-1})^{2}}\over{\lambda^{2}}}
\end{equation}
If Eve operates the teleporter with gain \cite{note3}
\begin{equation}
\lambda_{opt}={{1+V_{sq}^{2}}\over{1-V_{sq}^{2}}}
\end{equation}
where $V_{sq}=(\sqrt{G}-\sqrt{G-1})^{2}$, then Bob's noise penalty is
\begin{equation}
V_{B}^{\pm}(\lambda_{opt})={{1}\over{2 G-1}}
\end{equation}
and so Eve causes the minimum allowable disturbance, i.e. 
$V_{E}^{\pm}V_{B}^{\pm}=1$.

\section{Conclusion}
 
In this chapter we have investigated continuous variable quantum 
cryptography as it could be realized in optics by analysing the 
security and efficiency of specific implementations of two systems 
based on coherent and squeezed state light respectively. An Eve 
employing an optimal eavesdropper attack is assumed throughout. A 
possible optimal attack strategy that Eve could employ is outlined. 

We find that the coherent scheme can be made secure, but is not very 
efficient. None-the-less, given the maturity of optical communication 
technology based on rf modulation, this system may be competitive 
with discrete schemes in a local network scenario.

The squeezed state scheme can also be made secure and in principle is 
more efficient than the coherent state system. However its greater 
sensitivity to losses could make it less practical. 

We have looked at simple protocols throughout this analysis. It could be 
expected that more sophisticated encoding, reconciliation and 
privacy amplification techniques would lead to significant 
improvements in performance.

\section*{Acknowledgements}

We thank Michael Nielsen for useful discussions. This 
work was supported by the Australian Research Council.

%\begin{chapthebibliography}{1}

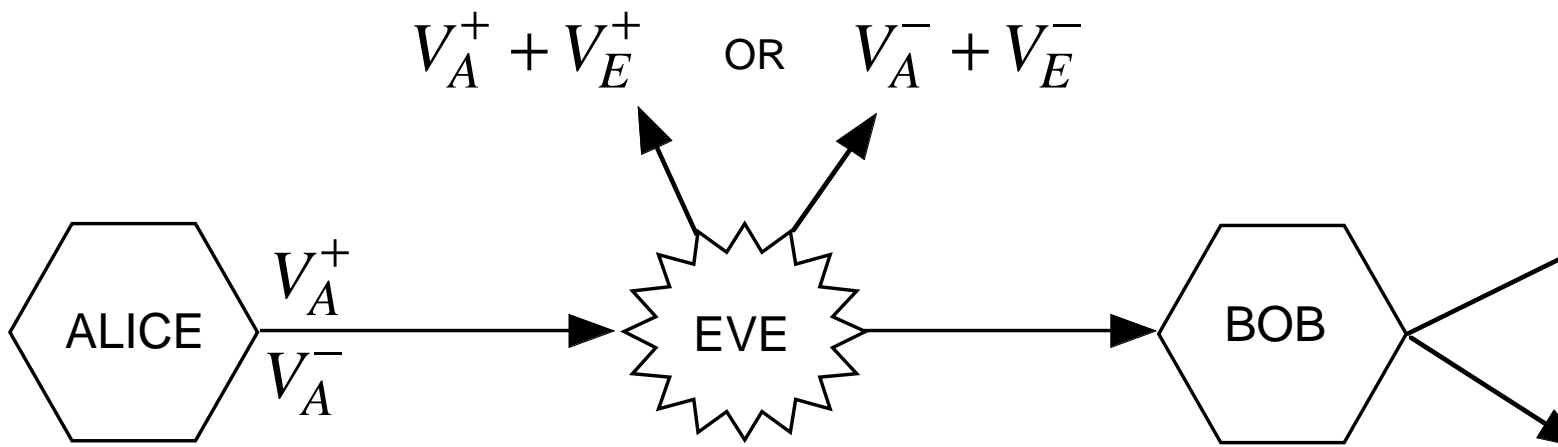
\begin{figure}[ht]
\caption{Schematic of general set-up. Alice sends information encoded 
in the amplitude ($V_{A}^{+}$) and phase ($V_{A}^{-}$) spectra. Bob 
makes measurements of either the amplitude or phase quadrature. Some 
additional noise is present on his measurements, $V_{B}^{\pm}$. Eve 
does not know which quadrature Bob will measure thus she needs to be able 
to extract information about both quadratures from her intercepted 
material. This leads to strict bound on the allowed values of the 
additional noise which must appear on her measurements ($V_{E}^{\pm}$)}
\end{figure}

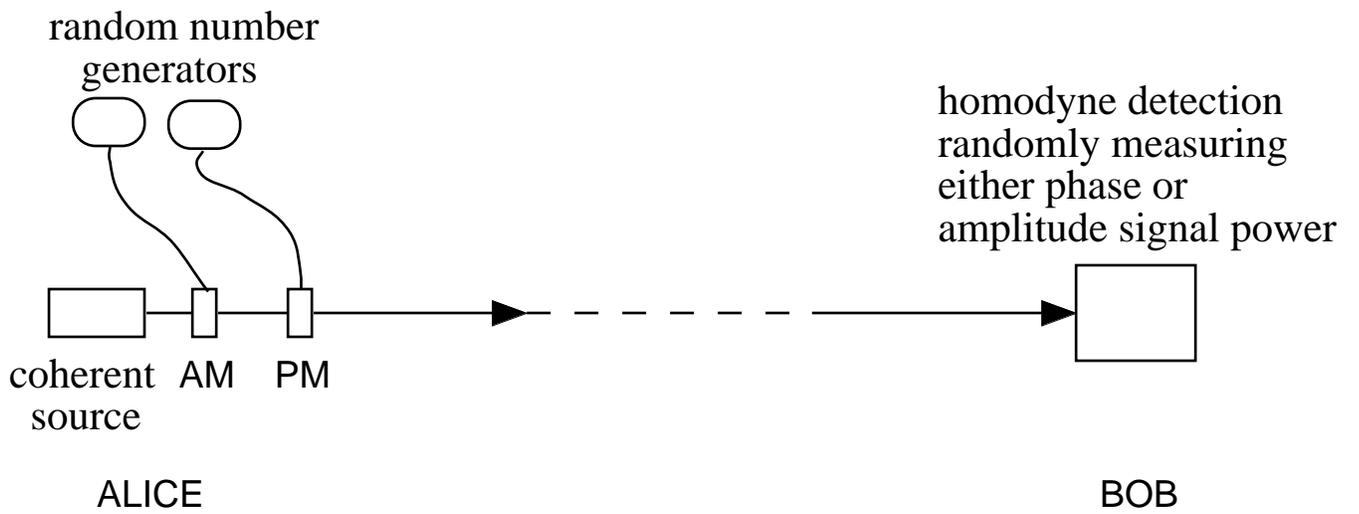
\begin{figure}[ht]
\caption{Schematic of coherent light cryptographic set-up. AM is an 
 amplitude modulator whilst PM is a phase modulator.}
\end{figure}

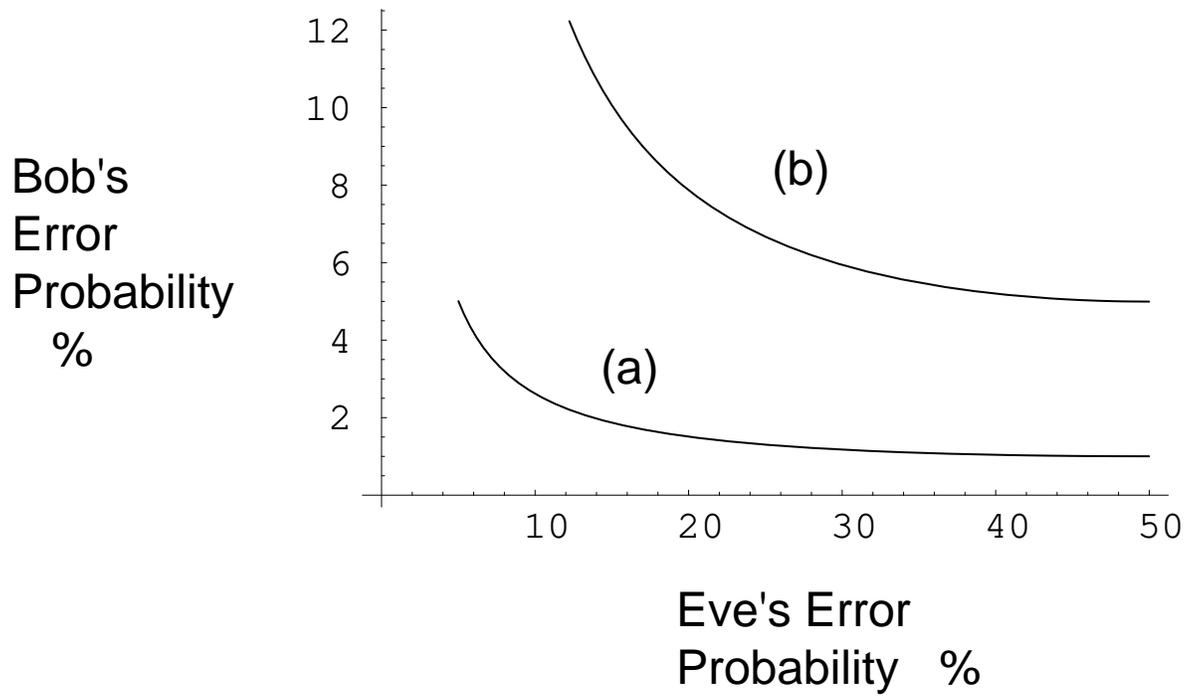
\begin{figure}[ht]
\caption{Minimum allowable error probabilities in the data of Bob and Eve 
are plotted for two signal to noise levels of Alice's beam. Trace (a) 
is for a signal to noise of 13dB whilst trace (b) is for a signal to 
noise of 10dB. }
\end{figure}

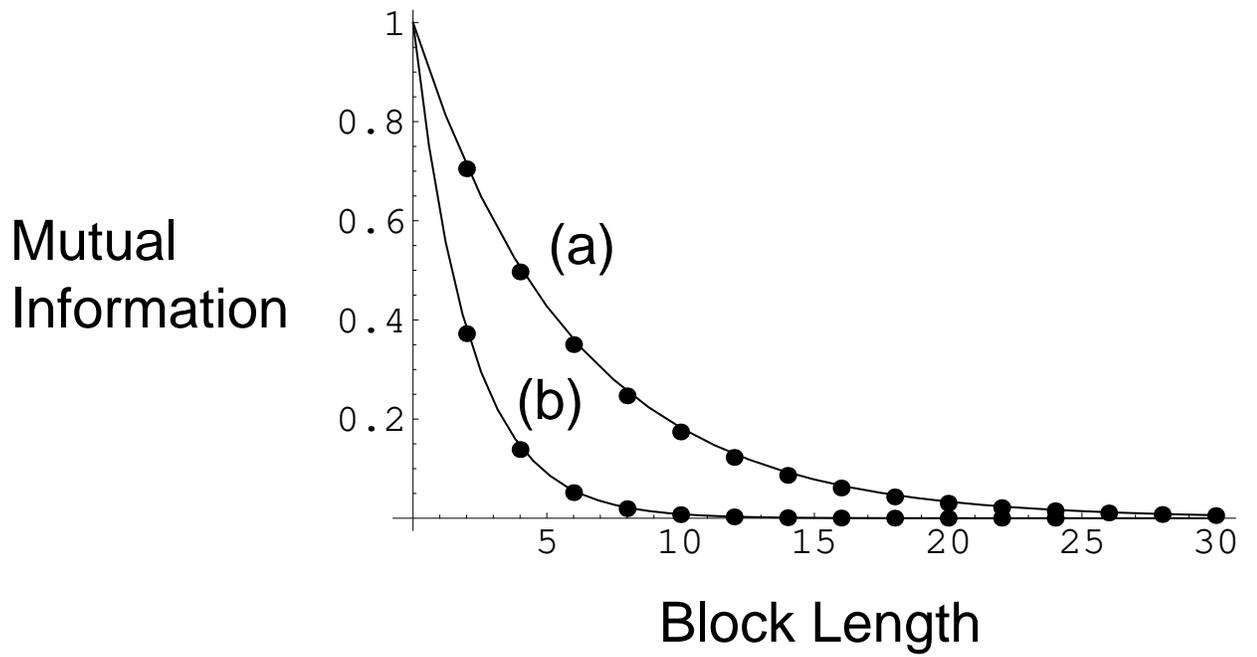
\begin{figure}[ht]
\caption{Decay of Eve mutual information as a function of the block 
length, $n$, in Alice and Bob's privacy amplification protocol
is plotted for two signal to noise levels of Alice's beam. Trace (a) 
is for a signal to noise of 13dB whilst trace (b) is for a signal to 
noise of 10dB. The solid traces are exponential fits.}
\end{figure}

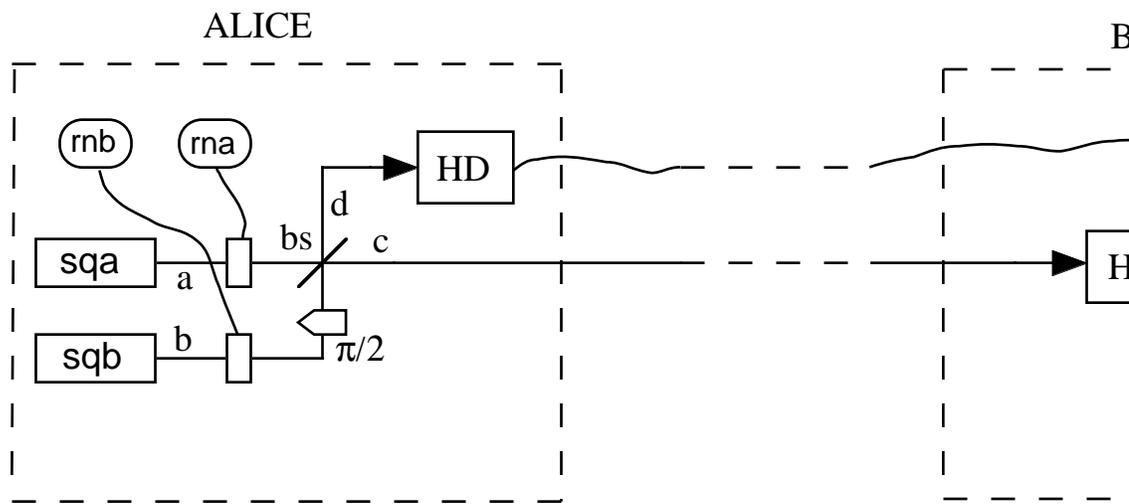
\begin{figure}[ht]
\caption{Schematic of squeezed light cryptographic set-up. Sqza and 
 sqzb are phase locked squeezed light sources. Rna and Rnb are 
 independent random number sources. Bs and pbs are non-polarizing and 
 polarizing beamsplitters respectively. Half-wave plates to rotate the 
 polarizations are indicated by $\lambda/2$ and optical amplification 
 by $A$. The $\pi/2$ phase shift 
 is also indicated. HD stands for homodyne detection system.}
\end{figure}

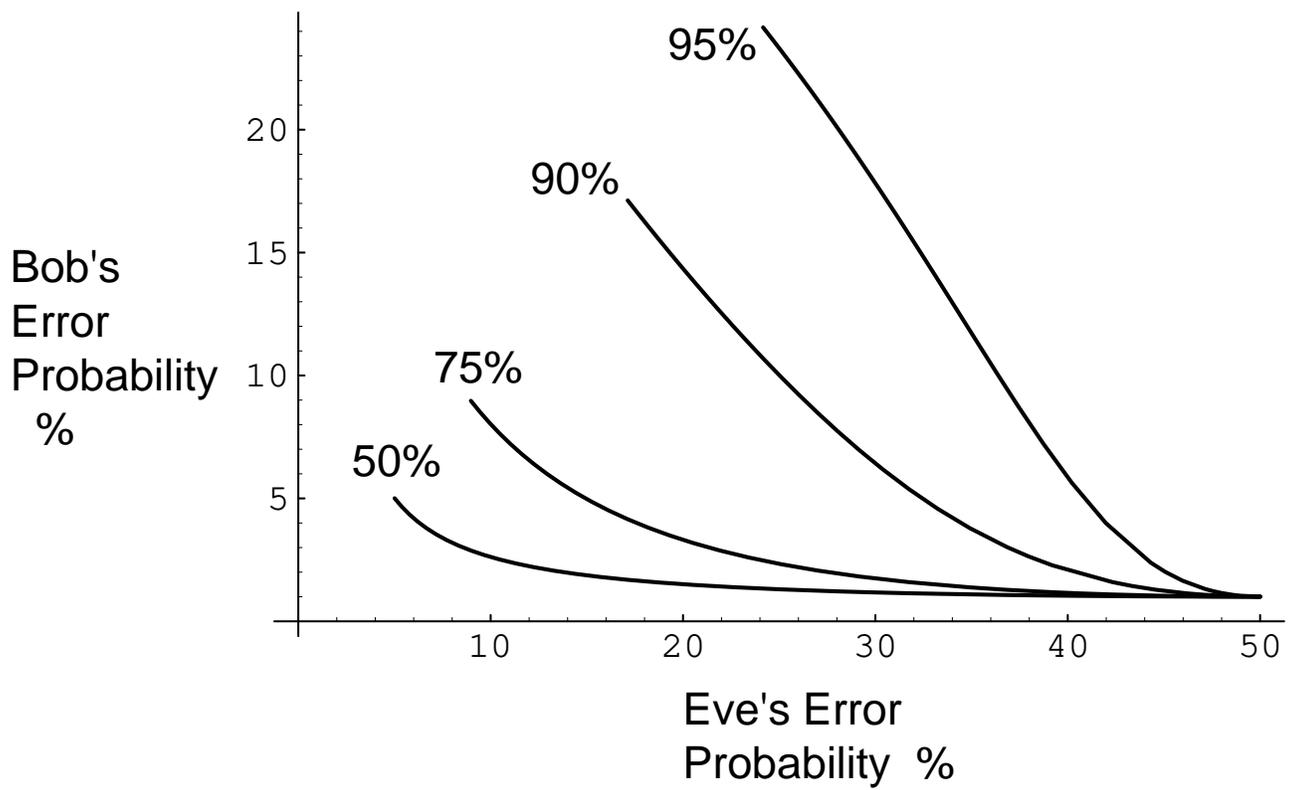
\begin{figure}[ht]
\caption{Minimum allowable error probabilities in the data of Bob and Eve 
are plotted for various levels of squeezing. }
\end{figure}

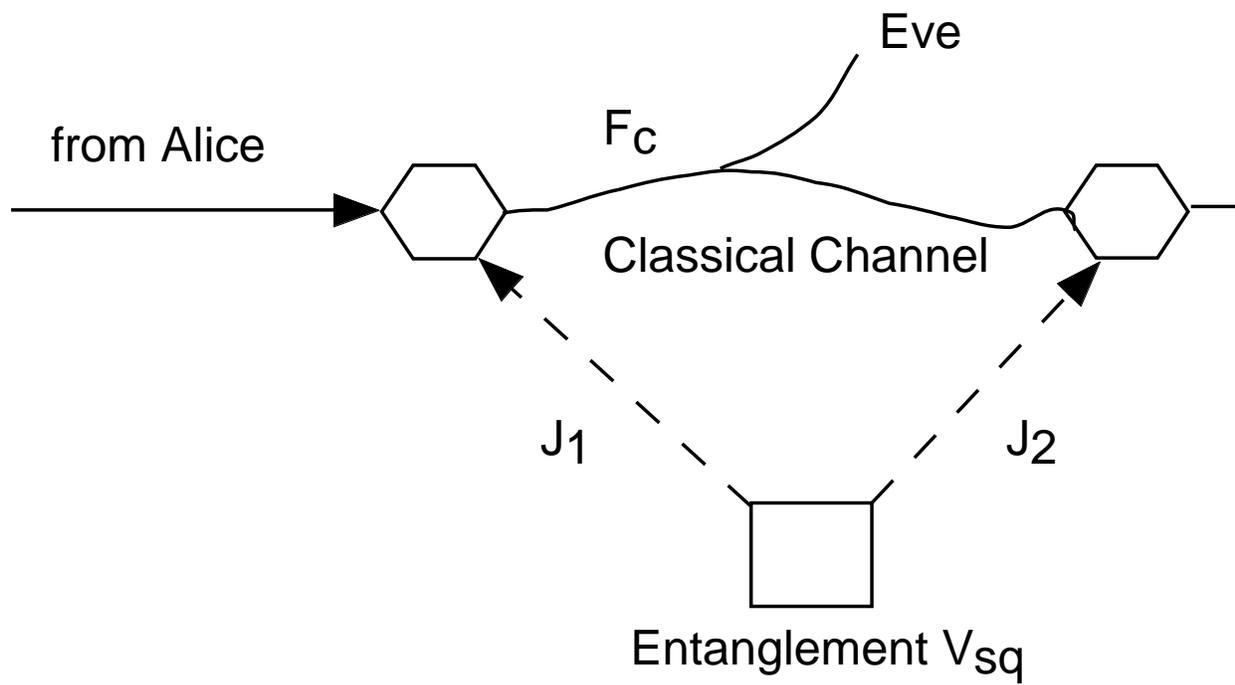
\begin{figure}[ht]
\caption{Schematic of teleportation being used as an optimum 
eavesdropper strategy. }
\end{figure}


\begin{thebibliography}{99}
\bibitem{note1} We only explicitly consider individual eavesdropper attacks here.
\bibitem{note2} This is a stronger assumption about Eve's capabilities 
than was used in assessing the effect of losses in Ref.[10].
\end{thebibliography}
\end{document}